\documentclass[journal]{IEEEtran}

\ifCLASSOPTIONcompsoc
\usepackage[caption=false, font=normalsize, labelfont=sf, textfont=sf]{subfig}
\else
\usepackage[caption=false, font=footnotesize]{subfig}
\fi

\usepackage{soul}
\usepackage{enumerate}
\usepackage[normalem]{ulem}
\usepackage{xcolor}
\usepackage{booktabs}

\newif\ifreview
\reviewtrue

\ifreview
\newcommand{\rv}[2][]{%
    \if\relax\detokenize{#1}\relax%
        \textcolor{blue}{#2}%
    \else%
        \textcolor{gray}{\sout{#1}}\textcolor{red}{#2}%
    \fi%
}
\newcommand{\todo}[1]{\noindent\textcolor{orange}{To-Do: {#1}}}
\newcommand{\comment}[2][]{
    \if\relax\detokenize{#1}\relax%
        \noindent\textcolor{magenta}{Comment: {#2}}%
    \else%
        \noindent\textcolor{magenta}{Comment from #1: {#2}}%
    \fi%
}
\else
\newcommand{\rv}[2][]{#2}
\newcommand{\todo}[1]{\empty}
\newcommand{\comment}[2][]{\empty}
\fi

\newcommand{\ie}{\textit{i.e.}}
\newcommand{\eg}{\textit{e.g.}}
\newcommand{\etal}{\textit{et~al.}}

\usepackage{graphicx}
\usepackage{multirow}

\begin{document}



\title{Deep Learning in Beyond 5G Networks with \\ Image-based Time-Series Representation}

\author{Lucas Fernando Alvarenga e Silva, Bruno Yuji Lino Kimura, and Jurandy Almeida\thanks{Lucas F. A. Silva, Bruno Y. L. Kimura, and Jurandy Almeida are with Institute of Science and Technology, Universidade Federal de S\~{a}o Paulo (UNIFESP).}}

\maketitle

\begin{abstract}
Towards the network innovation, the Beyond Five-Generation (B5G) networks envision the use of machine learning~(ML) methods to predict the network conditions and performance indicators in order to best make decisions and allocate resources. In this paper, we propose a new ML approach to accomplish predictions in B5G networks. Instead of handling the time-series in the network domain of values, we transform them into image thus allowing to apply advanced ML methods of Computer Vision field to reach better predictions in B5G networks. Particularly, we analyze different techniques to transform time-series of network measures into image representation, e.g., Recurrence Plots, Markov Transition Fields, and Gramian Angular Fields. Then, we apply deep neural networks  with convolutional layers to predict different 5G radio signal quality indicators. When comparing with other ML-based solutions, experimental results from 5G transmission datasets showed the feasibility and small prediction errors of the proposed approach.
\end{abstract}

\IEEEpeerreviewmaketitle

\section{Introduction}
\label{sec:introduction}

As an enabler for the future applications, the Fifth Generation Networks~(5G) have been deployed around the world. Essentially, the 5G is a triad of services to enable enhanced mobile broadband~(eMBB), the ultra-reliable low-latency communications~(uRLLC), and the massive machine-type communications~(mMTC). Network Function~(NF) allows network management and operation to be easily deployed and maintained in a cloud-based system, so-called the 5G Core. 

To assess and management the 5G links, the radio signal quality indicators are pivotal, \eg, Channel Quality Indicator~(CQI), Signal-to-Noise Ratio~(SNR), Reference Signal Received Quality~(RSRQ), Reference Signal Received Power~(RSRP), Received Signal Strength Indicator~(RSSI), among others. Such indicators are collected by the user equipment~(UE) and shared with the base station -- the evolved Node B~(eNB). Then, the eNB's radio network controller can adjust the channel modulation to ensure better communication links for the UEs. Since the indicator is an information of an event happened in the recent past, the channel modulations and other network management and operation are accomplished in a reactive manner. In 5G links, however, such reactive operation may not succeed, since the links are implemented with short-range high-frequency radio signal while the UEs are mobile nodes. To turn the management and operation proactive, the network information has to be predicted~\cite{DBLP:journals/corr/abs-2008-01000, DBLP:conf/iwmn/PareraRCLM19, DBLP:conf/wcnc/VankayalaS20, DBLP:journals/network/SakibTFFN21}.


Towards Beyond 5G Networks (B5G), accurate network information can be predicted from Machine Learning~(ML) models. While wide use of deep neural networks~(DNNs) with convolutional layers has been boosting scientific development in different fields, the Convolutional Neural Networks~(CNNs) becomes an important tool for computer networks. In this paper, we consider the recent advances in Computer Vision~\cite{DBLP:conf/icmv/HatamiGD17, DBLP:journals/prl/FariaAAMT16, DBLP:journals/lgrs/DiasDMLMT20} to best apply CNN to predict the signal quality indicators of radio link in B5G networks. More specifically, to perform regressions with CNN models, we propose to transform the one-dimensional network information time-series into images, \ie, a two-dimensional representation. When encoding the signal quality indicators (CQI, SNR, RSRQ, RSRP, RSSI) in images, the average magnitude of the errors observed in predictions  performed for 5G transmissions from file download and video streaming accomplished with static and vehicular nodes were of 2.4\% and 3.7\%, respectively. Compared with predictions of one-dimension time-series with LSTM (Long Short-Term Memory), which is a DL-model widely used for time series forecasting, the proposed approach could provide prediction almost four times more accurate on average.

The main contributions of our proposed approach are the following:
\begin{itemize}
    \item An evaluation of time-series of 5G radio signal quality signal indicators acquired from 83~datasets~\cite{DBLP:conf/mmsys/RacaLSQ20} of a mobile operator.
    \item Transformation of one-dimensional network information time-series into images by using different techniques, e.g., Recurrence Plots (RP), Markov Transition Fields (MTF), and Gramian Angular Fields (GAF). 
    \item The use of CNN, that are mainly used on computer vision tasks, to extract features and perform regression to predict the next future values for the 5G radio signal quality signal indicators.
\end{itemize}

The rest of this paper is organized as follows. We provide the background on deep learning and image-based time-series representation. We describe our ML approach and the experimental setup. Then, we discuss the experimental results under several performance aspects. Finally, we present our conclusions and future directions.

\section{Deep Learning}
\label{sec:DL}
As sub set of ML methods, Deep Learning~(DL) refers to a stack of layers of specific functions. The layers are coupled to jointly learn the ``rule of the game'' from data samples. It updates the internal weights in a backpropagation fashion, while ensuring the optimization of a given loss function by looking for its optimum throughout the variations of gradient analysis.

\subsection{Convolutional Neural Networks}

CNNs are a class of DL models that make use of a specific type of layer, so-called convolutional layer. In such layers, small filters extract local features from input data through the convolution operation, producing feature maps. When stacking the convolutional layers, a CNN model is able to learn hierarchically important features of a dataset, bringing expressive results in large and challenging benchmarks, like ImageNet~\cite{DBLP:journals/ijcv/RussakovskyDSKS15}.

\subsection{Loss Functions}

ML models are driven to objective functions, which translate the problem into a mathematical formula to be minimized or maximized. In regressions, a common objective function is the Mean Squared Error~(MSE), which is a loss function defined by a quadratic scoring method. Larger errors/outliers are penalized with higher weights, since the penalty is not proportional to the error but to its square. When minimizing the MSE, the ML model learns the (conditional) mean of the value. To achieve accurate predictions, however, it is important to measure the prediction uncertainty, while MSE itself may not be sufficient. Modelling uncertainty requires a suitable representation of the underlying probability distribution. Thus, a quantile loss function can replace a single value prediction by prediction intervals. 

\section{Image-based Time-Series}
\label{sec:time-series}

Time-series is a sequence of data presented in an temporal order. While the dynamics of the network transmissions are observed along the time, the order in which each data point is presented plays a key role. In the following, we present the transformation techniques we applied in our approach for encoding time-series into images.


\subsection{Recurrence Plots}
Recurrence plot (RP) is a data visualization technique applied to catch repetitions of events. In other words, it captures the periodic and/or cyclic behaviours in high dimensional data and projects them into two dimensional data, \ie, images. In its general formulation, it calculates a norm between two high-dimensional inputs, subtracts it from a threshold term and pass through a step function for each data point. In this work, we use a simplified RP as in~\cite{DBLP:journals/prl/FariaAAMT16}, specifically, a L1-norm zero-thresholded and one-dimensional input. The input time-series (as column vector) is repeated through the column dimension for a number of times equal to its length. Now, from a square matrix, it is subtracted by its transposed version and, finally, its absolute values are taken as the output. In this way, each column of the RP matrix represents the L1-distance between a point and all the other ones, allowing to visualize the variation of each data point in the time-series.

\begin{figure*}[h]
  \centering
  \includegraphics[width=\textwidth]{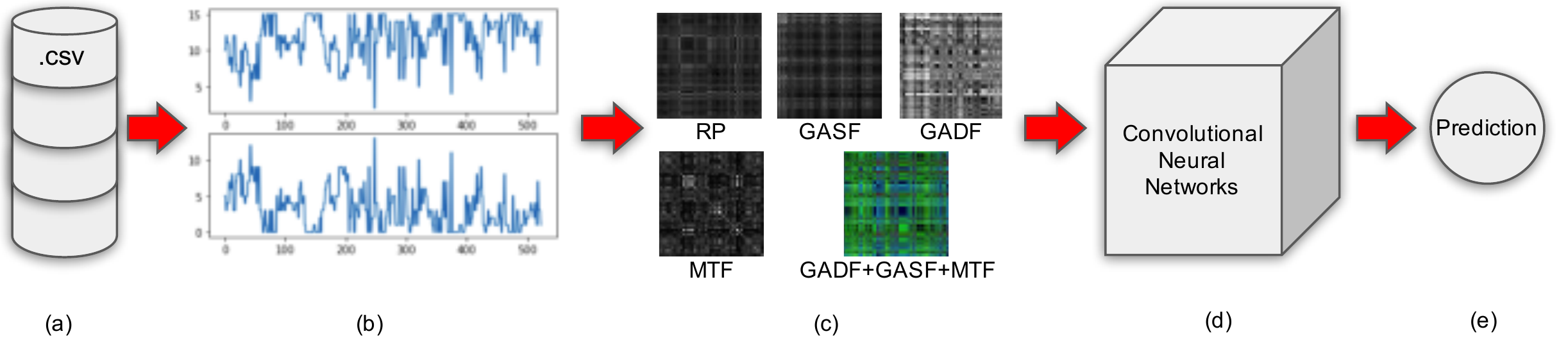}
  \caption{The data flow in the proposed approach: (a) datasets in ordinary format; (b) time-series with 5G network instance of interest; (c) time-series encoded in images using different techniques; (d) a CNN-model that learns the image extracts and predicts data points; and (e) data points predicted.}
  \label{fig:method}
\end{figure*}

\subsection{Gramian Angular Fields}

Gramian Angular Fields (GAF) derive from the Gramian matrix in linear algebra, which is obtained from the inner product between pairs of vectors in a set. In the usual application (Euclidean space), the inner product is determined from the scalar product between the vector's L2-norms and the cosine of the angle in between. However, if all the vectors have unit norm, the inner product can be simplified to the cosine of the angle between the pair of vectors. Then, the Gramian matrix can be calculated from the cosine of the angles. When calculating the inner product by the scalar product between the vectors, the GAF temporally correlates n-to-n time-series data points, \ie, a similarity measure between the vectors inherent to the scalar product.

In this paper, we use the Gramian Angular Summation Field (GASF) and Gramian Angular Difference Field (GADF) introduced in~\cite{DBLP:journals/lgrs/DiasDMLMT20}, which apply the trigonometrical sum and the trigonometrical difference between the angles of the vectors, respectively. To apply GASF and GADF transformation, we normalize the time-series to match the domain of the trigonometrical basic functions (sine and cosine), \ie, to the interval between -1 and 1, transform time series points in angles by applying the arc-cosine to all points. Finally, we construct the GASF matrix from the cosine of the sum of two angles, and the GADF matrix from the sine of the difference between two angles.

\subsection{Markov Transition Fields}

While RP carries the time-series variation information, the GAF carries the similarity between the time-series points. Differently, Markov Transition Fields (MTF) encode into an image the statistics of the transition between the states. Such an encoding is accomplished as follows:
\begin{enumerate}[(1)]
    \item Discretize the time-series into a defined number of bins in a uniform manner. Each bin refers to different intervals with the same length.
    \item Fill a square matrix, W, which  has the same dimensionality of the number of bins with the time-series transitions. It iterates through the time-series and increments the position in W that refers to that transition occurrence.
    \item Normalize each row of W in order to have the summation equal to one.
    \item From new square matrix, MTF, which has the same length of the time-series, fill MTF matrix with the values present in W according to the time-series transition states.
\end{enumerate}

\section{The Proposed Approach and Experimental Setup}
\label{sec:method}

The data flow of the proposed approach is shown in Fig.~\ref{fig:method}. The data flows through three main stages. First, from a dataset storing transmission logs in ordinary format (\eg, csv), we preprocess the dataset in order to extract the corresponding 1D times-series with 5G features of interest. In second stage, using the transformation techniques, the 1D times-series are then transformed into 2D representations, \ie, images. Finally, in the third stage, a CNN learns and predicts data points in the images. In the following, we present the experimental setup we implemented in each stage.

\subsection{The Dataset and Its Preprocessing}

We investigate the dataset presented in~\cite{DBLP:conf/mmsys/RacaLSQ20}. It has 83 different registers of Internet transmissions captured by an application (Android's G-NetTrack v18.7) in a smartphone (Samsung's S10) connected to a mobile operator in Irish. There are 3142 minutes of user-sided transmission logs that are organized into: three different services (\textit{File Download}, \textit{Amazon Prime}, and \textit{Netflix}); two different mobility patterns (\textit{Static} and \textit{Vehicular}); and, if the service refers to video streaming (\textit{Amazon Prime} or \textit{Netflix}), which content the user was consuming. The transmission logs were captured within the user's data plan limits (80GB), thus avoiding the downlink rate to be constrained and affect the measures when not exceeding the data plan limits. 

The transmission logs are stored in a csv file with a fixed number of features and a variable number of data points. The existing features are the following: timestamp; geographical coordinates; node velocity;  mobile operator (anonymized); cell id; network mode (5G, LTE, HSUPA, UMTS, HSPA+); downlink and uplink bitrates;  device state (idle or downloading); latency statistics (ping); signal quality indicators (SNR, RSRP, RSRQ, RSSI, CQI); and signal quality indicators of neighbours cells.

To construct the time-series, the transmission log is firstly preprocessed. Specifically, we sanitize the dataset by replacing Not-a-Number (NaN) entries with zero. Then, the data is sub-sampled to hold only the points of the 5G network mode. Finally, we select the features of interest, \ie, the signal quality indicators SNR, RSRP, RSRQ, RSSI, and CQI. Each of the 83 transmission logs is evaluated individually, that is, each one has its DL-model that is trained from scratch. 

\subsection{Time-series Transformation}

The time-series are iterated by a sliding window of size 32 and stride of 1. For each sliding window, the current data points are standardized into zero-mean and unit-deviation based on the statistics (\ie, mean and standard deviation) of previous data points. 
The idea is to simulate a real scenario where only past data points are known and, hence, it is not possible to compute the mean and standard deviation taking into account future data points.
Then, the standardized data points in the window are encoded into image through a transformation technique ({RP}, {GASF}, {GAFD}, {MTF}). If more than one technique is desired, a color image can be generated (\eg, {GADF+GAFD+MTF}).

\begin{figure*}[]
    \centering
    \subfloat[]{
        \includegraphics[width=0.31\textwidth]{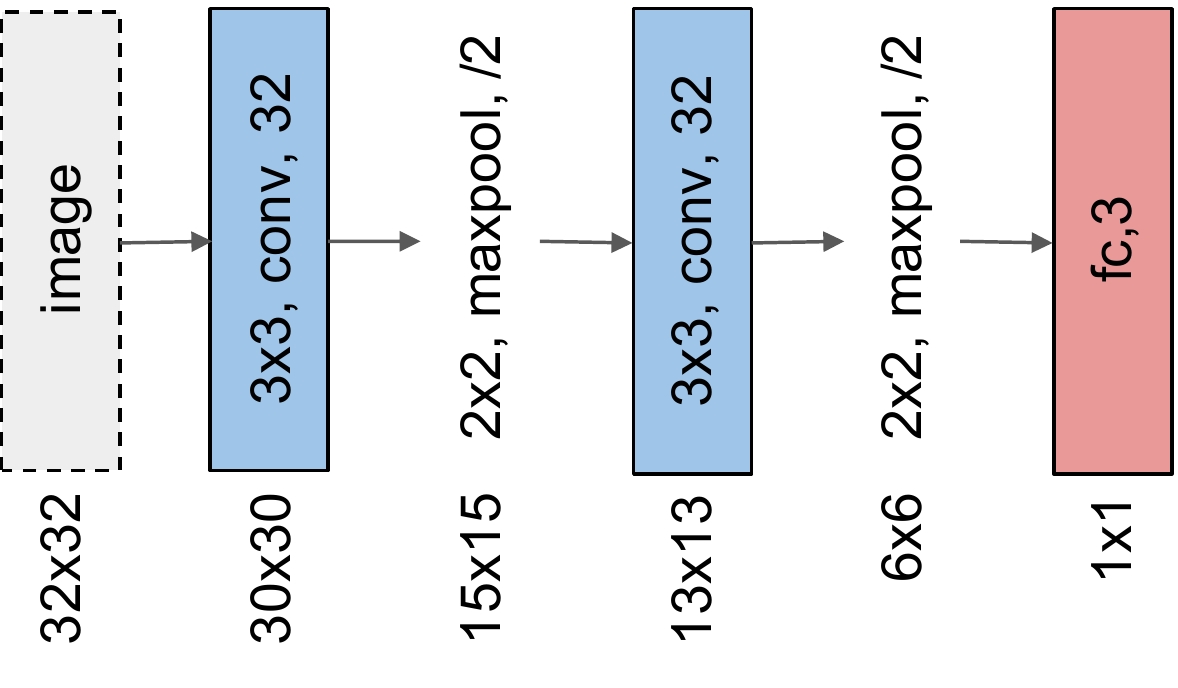}
        \label{fig:cnn:hatami}
    }
    \qquad
    \subfloat[]{
        \includegraphics[width=0.6\textwidth]{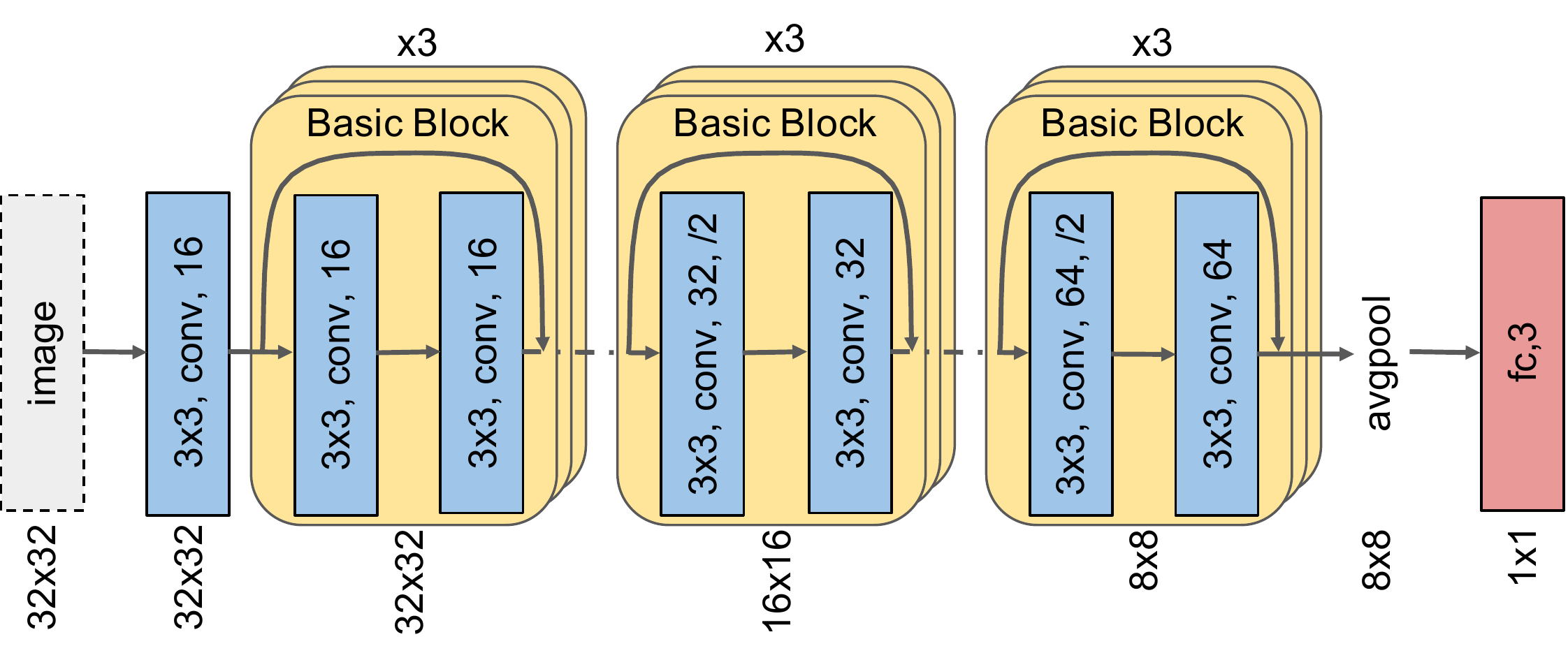}
        \label{fig:cnn:resnet}
    }    
    \caption{The DL models applied to extract features from the image-based time-series: (a) a shallow CNN introduced by Hatami~\etal~\cite{DBLP:conf/icmv/HatamiGD17}, and (b) a very deep CNN based on a variant of ResNet~\cite{DBLP:conf/cvpr/HeZRS16}, named ResNet-20.}
    \label{fig:cnn}
\end{figure*}

\begin{figure*}[t]
  \centering
  \includegraphics[width=\textwidth]{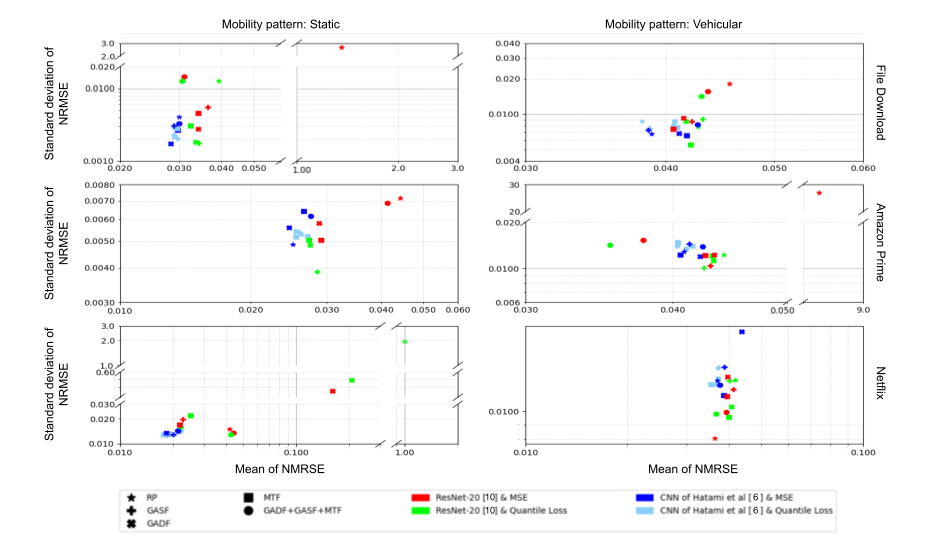}
  \caption{Distribution of the results obtained from the CQI prediction experiments. 
  }
  \label{fig:cqi-results}
\end{figure*}

\subsection{The Deep Learning Pipeline}

In our approach, we extract features from the image-based time-series by applying two different DL models, as illustrated in Fig.~\ref{fig:cnn}. The first one is a shallow CNN introduced by Hatami~\etal~\cite{DBLP:conf/icmv/HatamiGD17} (Fig.~\ref{fig:cnn:hatami}), which has only 3 layers with learnable parameters. On the other hand, the second model is a very deep CNN based on a variant of ResNet~\cite{DBLP:conf/cvpr/HeZRS16}, named ResNet-20, which has 20 layers with learnable parameters (Fig.~\ref{fig:cnn:resnet}). Besides popular CNN models, we choose these two ones due to their simplicity, allowing to meet performance requirements of B5G networks such as time constraints.

The set of generated images is divided into two subsets. The first one consists of 80\% of the samples, which is used as a training set. The remaining 20\% samples then belong to the test set. Each CNN-model is randomly initialized and trained to minimize the loss function (MSE or quantile loss) on the training set. During the training phase, the loss function determines the regression error to be back-propagated throughout the CNN in order to adjust its weights. 
We trained the CNNs from scratch for 120 epochs with a mini-batch size of $32$ by using the Adam optimization algorithm~\cite{DBLP:journals/corr/KingmaB14} with a weight decay of $5\times10^{-4}$ and an initial learning rate of $0.01$ with scheduled decay by a factor of 10 at the epochs 50 and 90. For the quantile loss, we changed the output layer to have 3 neurons representing the quantiles ($\gamma$) of $0.05$, $0.50$ and $0.95$. The $0.50$-quantile corresponds to the median and was taken as the model's prediction whereas the other two were used as a confidence interval of $90\%$ (\ie, from $0.05$ to $0.95$). Once the CNN-model is trained, in the test phase we accomplish the predictions, \ie, we take the test set as input for the CNN and analyse its output predictions.

\section{Experimental Evaluation}
\label{sec:results}


The MSE enlarges the absolute error by its square in order to penalize the misleading predictions. To analyse the error results in the same unit as the ground-truth, we calculate the square-root over MSE, \ie, RMSE. Then, we normalize RMSE (NRMSE) with the MinMax scaler within the groups of the same type of application service and mobility pattern. The NRMSE indicates an error value between 0 and 1, allowing the errors to be intuitively mapped into percentages. In the following, we discuss the NRMSE results regarding the robustness of the proposed method in predictions of different signal quality indicators, while comparing with other DL-model baselines. 

\subsection{Best prediction techniques}

Firstly, we focus on the CQI, since it plays a key for the network operators to best allocate channels. The mean and standard deviation of NRMSE results for CQI are shown in Fig.~\ref{fig:cqi-results}, regarding the image-based time-series transformation techniques and DL-models with different loss function. The more the points are close to zero on both axis, the better the prediction results are. Except for a few outliers, NRMSE results tends to zero in different combinations of techniques in the proposed approach. In the following, we discuss the experimental results.

In general, CNN of Hatami~\etal~\cite{DBLP:conf/icmv/HatamiGD17} performed better than ResNet-20~\cite{DBLP:conf/cvpr/HeZRS16}. By analyzing the results achieved by the CNN of Hatami~\etal~\cite{DBLP:conf/icmv/HatamiGD17}, we can notice that the performance of both loss functions (\ie, MSE and  quantile loss) was similar. However, the performance differences among the several settings evaluated for the combination of CNN of Hatami~\etal~\cite{DBLP:conf/icmv/HatamiGD17} and quantile loss function are the smallest in relation to all the other combinations of DL model and loss function. In other words, this means that the combination of CNN of Hatami~\etal~\cite{DBLP:conf/icmv/HatamiGD17} and quantile loss function is less sensitive to changes in the application services, mobility patterns, and time-series transformation techniques. For this reason, this  combination was chosen for the next experiments. 
Among the time-series transformation techniques tested jointly with CNN of Hatami~\etal~\cite{DBLP:conf/icmv/HatamiGD17} and quantile loss function, MTF yielded the lowest NRMSE for most of cases and, for this reason, it was chosen for comparison with other baselines.

\begin{figure*}[]
  \centering
  \includegraphics[width=\textwidth]{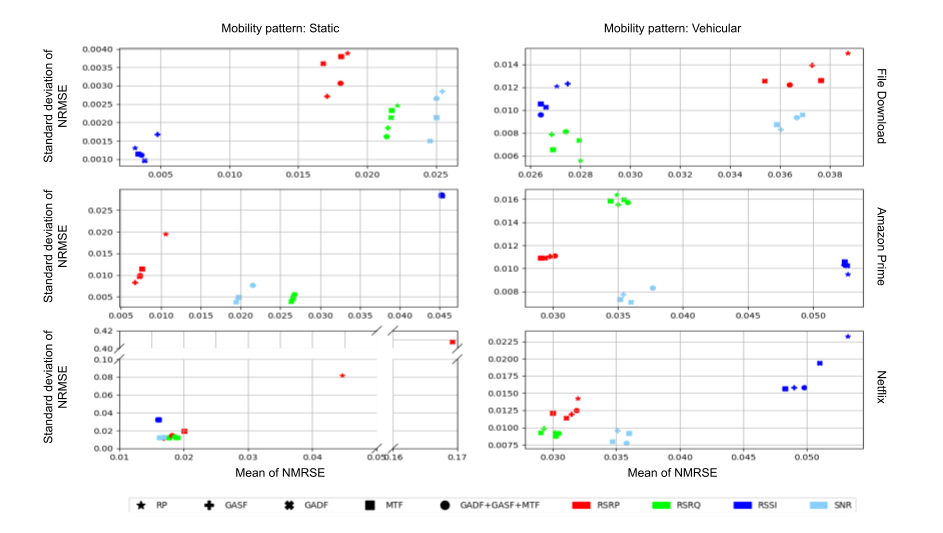}
  \caption{Distribution of the results of our approach, with CNN of Hatami~\etal~\cite{DBLP:conf/icmv/HatamiGD17} and minimization of quantile loss, to predict the signal quality indicators (SNR, RSRP, RSRQ and RSSI) of long range values.
  }
    \label{fig:other_features_results}
\end{figure*}

\subsection{Node mobility pattern}

From the point of view of the node mobility pattern, transmission from static nodes allowed better CQI predictions than vehicular ones, regardless of the application service (Download, Amazon Prime, and Netflix). This result is expected, since the signal indicators for static nodes are less fluctuating than those for mobile nodes, mostly vehicular nodes. Particularly, the wireless transmissions between a static UE and eNB are less exposed to the impacts of the fast-time varying channels (e.g., doppler effect, and fast fading) as in vehicular nodes. 

\subsection{Application services}

The applications have different patterns, specifically, the transmission workloads are significantly different. While a single control loop (TCP congestion control) determines the transmission rate of the download application, there are two control loops in the video applications (video adaptive bitrate, and the congestion control). As a result, when downloading a large file (200MB) via a greedy TCP connection over the 5G network, the UE could obtain a greater transmission rate (static 66.9 Mbps, vehicular 28.5 Mbps) than Netflix (static 13.7 Mbps, vehicular 7.5Mbps) and Amazon Prime (static 6.9 Mbps, vehicular 1.3 Mbps) applications. However, the different application transmission patterns had not impacted significantly the prediction results, as shown in Fig.~\ref{fig:cqi-results}. In this case, the predicted signal quality indicators are metrics of the physical layer, i.e., they are not dependent on  the type of application transmission behaviour.

\subsection{Robustness with other signal quality indicators}

With the best combined techniques in our approach (MTF, CNN  Hatami~\etal~\cite{DBLP:conf/icmv/HatamiGD17}, and quantile loss function) that we observed in CQI experiments, we then predict the other indicators of long range values (SNR, RSRP, RSRQ, and RSSI). The distribution of means and standard deviations of NRMSE results are presented in Fig.~\ref{fig:other_features_results}. As can be seen, the results are clustered around the signal quality indicator, while not being so close to zero as in CQI predictions. However, the major prediction errors are less than 10\%. When being clustered, all the different image-based transformation techniques are near from each other, showing that they behave similarly when predicting the indicators of range value longer than CQI.



\subsection{Baseline Comparison}

To establish baseline results to compare, we evaluate the one-dimensional CNN (1D-CNN) and the LSTM approach proposed by Parera~\etal~\cite{DBLP:conf/iwmn/PareraRCLM19} to predict CQI on 4G LTE networks schemes. Particularly, our work differs by two major aspects: (1) we are dealing static and vehicular transmission in B5G networks, a challenger scenario due to the high dynamic pattern of short-range high-frequency radio signals; and (2) our samples are spaced in one second apart, while their approach regards time-series consisted of values summarized one hour apart, e.g, one-hour average of the CQI. Using the same 5G datsets~\cite{DBLP:conf/mmsys/RacaLSQ20} as in our experiments, we evaluate the baselines in four ways: (\textit{i}) the 1D-CNN with its training hyper-parameters, i.e., batch-size of 128 samples, 300 epochs with early-stopping, optimizer Adam~\cite{DBLP:journals/corr/KingmaB14}, learning rate of $0.001$; (\textit{ii}) the statefull LSTM with its training hyper-parameters, the same as in (\textit{i}) but without early-stopping and batch-size of 1 sample.


The observed average of NRMSEs are shown in Table~\ref{tab:baseline_results}. As can be seen, our proposed approach overcomes LSTM, while providing  predictions with errors slightly higher than 1D-CNN. It is important to notice that the 1D-CNN has much more \#Params than the 2D-CNN based on Hatami~\etal~\cite{DBLP:conf/icmv/HatamiGD17} (almost 29x more), as presented in Table~\ref{tab:model_metrics}, which contains the number of MACs (Multiply-ACcumulate operations) and Parameters of CNN models. In other words, the difference of prediction errors is negligible between the proposed approach and 1D-CNN, while our approach is much less costly. 


\begin{table*}[t]
\caption{Comparison of our approach with other baselines}
\label{tab:baseline_results}
\resizebox{\textwidth}{!}{%
\begin{tabular}{c|c|c|c|c|c|c|c}
                                          \toprule \multirow{2}{*}{Approach} & \multicolumn{2}{c|}{Download}                              & \multicolumn{2}{c|}{Amazon Prime}                          & \multicolumn{2}{c|}{Netflix}                               & \multirow{2}{*}{Average}                       \\ \cmidrule{2-7}
                                          & Static                      & Dynamic                     & Static                      & Dynamic                     & Static                      & Dynamic                     &                            \\ \midrule
CNN-1D of Parera~\etal~\cite{DBLP:conf/iwmn/PareraRCLM19}            & 0.0192 $\pm$ 0.00418          & 0.0224 $\pm$ 0.00380          & 0.0177 $\pm$ 0.00540          & 0.0244 $\pm$ 0.00468          & 0.00176 $\pm$ 0.00334          & 0.0225 $\pm$ 0.00463          & 0.0217 $\pm$ 0.00495          \\ \midrule
LSTM of Parera~\etal~\cite{DBLP:conf/iwmn/PareraRCLM19}              & 0.105 $\pm$ 0.0340                       & 0.141 $\pm$ 0.0238                       & 0.0521 $\pm$ 0.0559                       & 0.136 $\pm$ 0.0380                       & 0.0937 $\pm$ 0.0273                       & 0.141 $\pm$ 0.0360                       & 0.123 $\pm$ 0.0451                       \\ \midrule
MTF + CNN of Hatami~\etal~\cite{DBLP:conf/icmv/HatamiGD17} (ours)             & 0.0276 $\pm$ 0.00293          & 0.0409 $\pm$ 0.00800           & 0.0165 $\pm$ 0.0128           & 0.0355 $\pm$ 0.0127           & 0.0255 $\pm$ 0.00547          & 0.0404 $\pm$ 0.0151           & 0.0343 $\pm$ 0.0139          \\ \bottomrule
\end{tabular}%
}
\end{table*}

\begin{table}[h]
\centering
\caption{\#MACs and \#Params of used CNN methods}
\label{tab:model_metrics}
\footnotesize
\begin{tabular}{l c c}
    \toprule
    DL Models & \#MACs(M) & \#Params(k) \\
    \midrule 
CNN-1D of Parera~\etal~\cite{DBLP:conf/iwmn/PareraRCLM19} & 9.48          & 396.8   \\ 
LSTM of Parera~\etal~\cite{DBLP:conf/iwmn/PareraRCLM19} & 0.0706        & 2.38            \\ 
CNN of Hatami~\etal~\cite{DBLP:conf/icmv/HatamiGD17}      & 2.37          & 13.6            \\ 
ResNet20~\cite{DBLP:conf/cvpr/HeZRS16}                    & 40.9          & 269.2        \\ 
\bottomrule 
\end{tabular}%
\end{table}

\section{Conclusions and Future Work}
\label{sec:conclusion}

Recently, many research efforts on the design, adaptation and enhancement of B5G networks have been directed towards developing smart networking with better efficiency for network management and operation. In this scenario, ML techniques are a key ally for better prediction and decision-making in resource allocation. In this paper, we have presented a novel DL approach to predict 5G radio signal quality indicators. Different from previous works, we transform network information time-series into images, enabling us to take advantage of CNNs extremely successful in computer vision applications. Our experiments analyzed five image-based time-series representations as input to fed two different CNNs designed to predict different 5G signal quality indicators. We validated our approach on 5G transmission data collected for different application services and mobility patterns. The obtained results showed that the proposed method is effective, achieving a NRMSE of 0.0343 $\pm$ 0.0139 on average.



In a future perspective, a possible next step includes deploying the proposed approach as NF and evaluate its performance gains in terms of Quality of Services (QoS) in B5G transmissions. Improvements on the steps of time-series transformation and best tuning DL-model for 5G communications also require further investigations.

\section*{Acknowledgements}
This research was supported by the FAPESP-Microsoft Research Virtual Institute (grants~2017/25908-6 and 2020/08770-3) and the Brazilian National Council for Scientific and Technological Development - CNPq (grant~314868/2020-8).

\bibliographystyle{IEEEtran}
\bibliography{paper}

\section*{Biographies}

\vskip -20pt plus -1fil

\begin{IEEEbiographynophoto}
	{Lucas Fernando Alvarenga e Silva} is with the Federal University of São Paulo at the Institute of Science and Technology, Brazil. He is M.Sc. candidate, whose research interest is in machine learning for computer vision, mainly based on convolutional neural networks for action recognition, object classification, and domain adaptation.
\end{IEEEbiographynophoto}

\vskip -20pt plus -1fil

\begin{IEEEbiographynophoto}
	{Bruno Yuji Lino Kimura} is an Associate Professor with the Federal University of São Paulo at the Institute of Science and Technology, Brazil. Currently, his research interests include performance improvement in multi-path communications, IoT/Edge networks, dependable microservice architectures, and machine learning applied to networks.
\end{IEEEbiographynophoto}

\vskip -20pt plus -1fil

\begin{IEEEbiographynophoto}
	{Jurandy Gomes de Almeida} is an Associate Professor with the Federal University of São Paulo at the Institute of Science and Technology, Brazil. 
	He has developed research on databases, image processing, machine learning, and computer vision in applications of visual information retrieval.
\end{IEEEbiographynophoto}

\end{document}